\documentclass[aps,onecolumn,showpacs,amsmath,nofootinbib]{revtex4-1}

\newcommand{\beqa}{\begin{eqnarray}}
\newcommand{\eeqa}{\end{eqnarray}}

\usepackage{graphicx}
\usepackage{epsfig}

\usepackage{graphicx}
\usepackage{dcolumn}
\usepackage{bm}%

\begin{document}

\title{The abundance of primordial black holes from the global 21cm signal and extragalactic gamma-ray background}

\author{Yupeng Yang$^{1,2}$}

\affiliation{$^1$School of Physics and Physical Engineering, Qufu Normal University, Qufu, Shandong, 273165, China\\
$^2$Joint Center for Particle, Nuclear Physics and Cosmology, Nanjing, Jiangsu, 210093, China}

\begin{abstract}
Primordial black holes (PBHs) formed in the early Universe can accret dark matter particles due to gravity and form ultracompact minihalos (UCMHs). The theoretical researches and simulations have shown that the density profile of dark matter in UCMHs is 
in the form of $\rho (r) \sim r^{-2.25}$. Compared with the popular dark matter 
halo model, e.g. NFW model, dark matter annihilation rate is larger in UCMHs. 
Considering dark matter annihilation, there is a maximum core density $\rho_{\rm max}$ in UCMHs which has been 
treated independent on redshift. While in this work, we point out that $\rho_{\rm max}$ depends on redshift and 
dark matter annihilation rate in UCMHs also changes with time. 
We re-investigate the $\gamma$-ray flux from UCMHs due to dark matter annihilation and focus on their contributions to 
the extragalactic gamma-ray background (EGB). 
Utilizing the EGB data of Fermi, the constraints on the abundance of PBHs are derived, $\Omega_{\rm PBH,EGB} \lesssim 2\times 10^{-8}$. 
Motivated by the recent observations of global 21cm signals by the EDGES experiment, 
we investigate the influences of dark matter annihilation in UCMHs on global 21cm signals 
and derive the constraints on the abundance of PBHs, $\Omega_{\rm PBH,21cm} \lesssim 2\times 10^{-10}$. 
The derived constraints are valid for mass range 
of $10^{-6} M_{\odot} \lesssim M_{\rm PBH} \lesssim 10^{3} M_{\odot}$. 

\end{abstract}
\maketitle

\section{introduction} 
Primordial black holes (PBHs), formed through the collapse of large 
density perturbations in the early Universe can accrete dark matter particles by gravity and 
gradually develop ultracompact minihalos (UCMHs), dressed PBHs. 
The density profile of dark matter in UCMHs is in the form of 
$\rho(r) \sim r^{-2.25}$, which have been proved by theoretical researches 
and simulations~\cite{Bertschinger,Ricotti:2007jk,0908.0735,
Adamek:2019gns,Boucenna:2017ghj,Eroshenko:2016yve}. 
Dark matter particles can annihilate into standard model 
particles shown by Weakly Interacting Massive Particles model~\cite{Jungman:1995df,Bertone:2004pz}, 
a popular researched dark matter model.  
It is expected that the annihilation rate of 
dark matter particles in UCMHs is larger than that 
of the classical dark matter halo model, e.g. NFW model. Therefore, UCMHs should have significant influences 
on the astrophysical observations and we will focus on the global 21cm signals and the extragalactic gamma-ray background (EGB).

Dark matter annihilation in UCMHs contributes to the EGB, which has been investigated by e.g. 
Refs.~\cite{Boucenna:2017ghj,Adamek:2019gns}, 
and the constraints on the abundance of PBHs are obtained. The basic idea of these works 
is that the contributions of dark matter annihilation in UCMHs to the EGB are similar to the case of 
dark matter decay. Therefore, using the relevant results of dark matter decay from the EGB data one can easily 
get the constraints on the abundance of PBHs. However, dark matter annihilation rate in UCMHs depends 
on the maximum core density ($\rho_{\rm max}$) and corresponding radius ($r_{\rm cut}$). 
In Refs.~\cite{Boucenna:2017ghj,Adamek:2019gns}, 
the authors treated $\rho_{\rm max}$ and $r_{\rm cut}$ independent on redshift. In this work, 
we stress that $\rho_{\rm max}$ and $r_{\rm cut}$ evolve with time. Therefore, it should not be analogous to the 
dark matter decay case directly. Using the complete expression of $\rho_{\rm max}(z)$and $r_{\rm cut}(z)$, 
we reinvestigate the contributions of $\gamma$-ray flux from UCMHs due to dark matter annihilation to 
the extragalactic gamma-ray background. Utilizing the EGB data of Fermi, we derive the 
constraints on the abundance of PBHs, $\Omega_{\rm PBH,EGB} \lesssim 2\times 10^{-8}$. 

Recently, the EDGES experiment has found the observational results of the global 21cm signals 
at the redshift of $z\sim 17$~\cite{edges-nature}. The amplitude of absorption trough in the spectrum 
is about $T_{\rm 21} = -500~\rm mK$, which is nearly twice the standard value. 
Many works have appeared for explaining the large absorption trough, e.g. Refs~\cite{Barkana:2018lgd,Feng:2018rje,prd-edges}. 
Based on the results of the EDGES experiment, the basic qualities of dark matter 
can be constrained, by requiring the absorption trough in global 21cm signals not erased 
by extra-sources~\cite{yinzhema,prd-edges,DAmico:2018sxd,Kovetz:2018zes,Bhatt:2019qbq,Berlin:2018sjs,Barkana:2018cct}. 
In this work, we will investigate 
the influences of dark matter annihilation in UCMHs on global 21cm signals. 
By requiring the bright temperature $T_{\rm 21} \lesssim -100~\rm mK$, we derive the constraints on the 
abundance of PBHs, $\Omega_{\rm PBH,21cm} \lesssim 2\times 10^{-10}$.

This paper is organized as follows: In Sec. II, the basic properties of UCMHs is discussed. 
In Sec. III, the contributions of dark matter annihilation in UCMHs on 
global 21cm signals and the EGB is investigated and then the constraints on the abundance of PBHs are derived, 
utilizing the EGB data by Fermi experiment and 21cm data by the EDGES experiment, 
we derive the constraints on the abundance of PBHs. The conclusions and discussions are given in Sec. IV.


\section{The basic properties of dressed PBHs}

In this section, we briefly review the basic properties of UCMHs (dressed PBHs) and 
one can refer to e.g. Refs.~\cite{Ricotti:2007jk,0908.0735,
Adamek:2019gns,Boucenna:2017ghj,Eroshenko:2016yve} for more detailed discussions. 
 
After formation of PBHs, dark matter particles can be accreted onto PBHs forming UCMHs. 
The evolution of the mass and radius of UCMHs is in the form as~\cite{Ricotti:2007jk,0908.0735,
Adamek:2019gns,Boucenna:2017ghj,Eroshenko:2016yve}  

\beqa
M_{\rm UCMH} = 3\left(\frac{1000}{1+z}\right)M_{\rm PBH} \\
R_{\rm UCMH} = 0.019~{\rm pc}\left(\frac{M_{\rm UCMH}}{M_{\odot}}\right)^{\frac{1}{3}}\left(\frac{1000}{1+z}\right)
\label{eq:mass_r_ucmh}
\eeqa

According to the theory and simulation, the density profile of dark matter within UCMHs is in the form of 
$\rho(r) \sim r^{-2.25}$~\cite{Bertschinger,Ricotti:2007jk,0908.0735,
Adamek:2019gns,Boucenna:2017ghj,Eroshenko:2016yve}. In this work, we adopt the form as~\cite{Eroshenko:2016yve,Adamek:2019gns}

\beqa
\rho(r) = 9.5\times 10^{-23} \left(\frac{r}{\rm pc}\right)^{-\frac{9}{4}}\left(\frac{\rm M_{\rm PBH}}
{\rm M_{\odot}}\right)^{\frac{3}{4}}~\rm g~cm^{-3}.
\label{eq:rho}
\eeqa 

As shown in Eq.~(\ref{eq:rho}), $\rho(r) \to \infty$ for $r\to 0$. Considering dark matter annihilation, 
there is a maximum core density $\rho_{\rm max}$ within UCMHs which 
can be written as~\cite{Bringmann_1,epjplus-2,dongzhang}

\beqa
\rho_{\rm max} = \frac{m_{\chi}}{\left<\sigma v\right>\left(t-t_{i}\right)},
\label{eq:rho_max}
\eeqa
where $m_{\chi}$ is the mass of dark matter particle, $\left<\sigma v\right>$ is a thermally averaged 
cross section. $t$ is the age of the Universe and $t_i$ is the formation time of UCMHs. 
For matter dominated epoch of the Universe, as a good approximation, the age of the Universe can be written as~\cite{dongzhang} 

\beqa
t(z) = \frac{2}{3}(1+z)^{-3/2}\Omega_{m,0}^{-1/2}H_{0}^{-1}. 
\label{eq:t}
\eeqa
Since we consider the effects of UCMHs in low redshift of $z\lesssim 1000$, therefore, 
$t(z) \gg t_i$ and $\rho_{\rm max}$ can be rewritten as
\beqa
\rho_{\rm max}(z) =  \frac{m_{\chi}}{\left<\sigma v\right>t(z)},
\eeqa

\begin{figure}
\epsfig{file=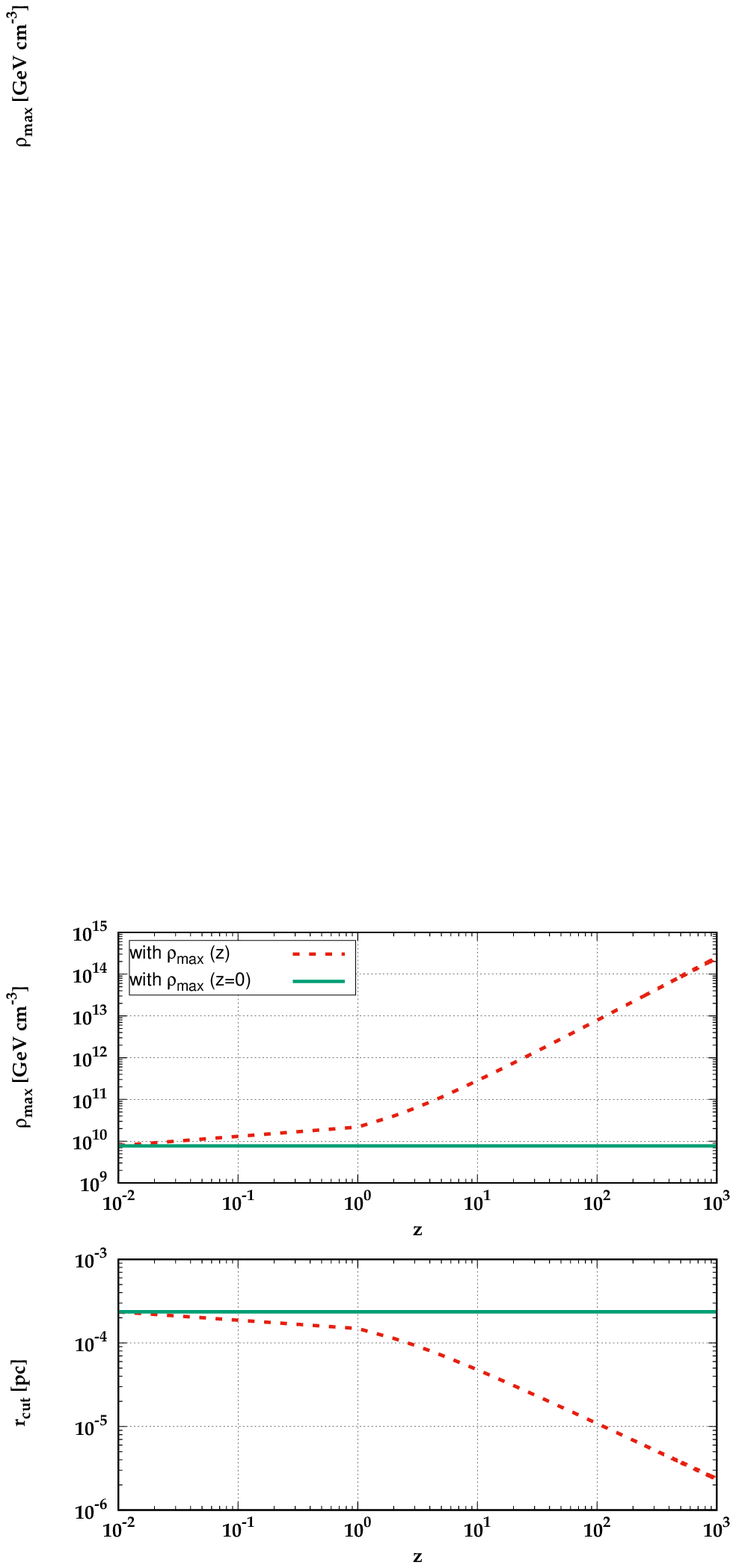,width=0.85\textwidth}
\caption{The evolution of $\rho_{\rm max}(z)$ and $r_{\rm cut}(z)$ (red dashed line). For comparison, 
we also show the values of $\rho_{\rm max}$ and $r_{\rm cut}$ at the redshift of $z=0$ (green solid line). 
We have set the parameters of dark matter as 
$m_{\chi} =100~\rm GeV$ and $\left<\sigma v\right>=3\times 10^{-26}\rm cm^{3}s^{-1}$. 
Here the mass of PBH is set as $M_{\rm PBH} = 1 M_{\odot}$.}
\label{fig:rho_r}
\end{figure}

In previous works, the authors of e.g.~\cite{Boucenna:2017ghj,Adamek:2019gns} have used $t(z=0)$ for their calculations 
and for this case $\rho_{\rm max}$ is independent on redshift. 
Due to the presence of $\rho_{\rm max}$, there is a cut of the radius $r_{\rm cut}$ and it can be determined by 
$\rho_{\rm max} = \rho(r_{\rm cut})$, 

\beqa
r_{\rm cut}= 1.6\times 10^{-10} {\rm pc}\left(\frac{m_{\rm \chi}}{\left<\sigma v\right>}\right)^{-\frac{4}{9}}
\left(\frac{M_{\rm PBH}}{M_{\odot}}\right)^{\frac{1}{3}}t(z)^{\frac{4}{9}}
\eeqa
Therefore, $r_{\rm cut}$ also depends on redshift. The evolution of $\rho_{\rm max}(z)$ and $r_{\rm cut}(z)$ 
is shown in Fig.~\ref{fig:rho_r} (red dashed line). For comparison, 
we also show the values of $\rho_{\rm max}$ and $r_{\rm cut}$ at the redshift of $z=0$ (green solid line), 
which have been used in e.g. Refs.~\cite{Boucenna:2017ghj,Adamek:2019gns}. 


Therefore, considering dark matter annihilation, the density profile of dark matter in UCMHs can be 
written as

\beqa 
\rho(r)=\left\{
             \begin{array}{lr}
		\rho_{\rm max},~~r\leq r_{\rm cut} &  \\
             \rho(r), ~~r_{\rm cut}<r\leq R_{\rm UCMH} &  
             \end{array}
\right.
\eeqa


\section{Constraints on PBHs from the EGB and global 21cm signals}

\subsection{Constraints on PBHs from the EGB data}

As mentioned above, the dark matter annihilation within UCMHs can contribute to the EGB. 
The corresponding $\gamma$-ray flux from UCMHs can be written as~\cite{yang_2011,Ullio:2002pj,Bergstrom:2001jj}

\beqa
\frac{d\phi_{\gamma}}{dE} = \frac{c}{8\pi}\frac{\Omega_{\rm PBH}\rho_{c,0}}{M_{\rm PBH}}\frac{\left<\sigma v\right>}{m_{\chi}^{2}}
\int dz\frac{e^{-\tau(E,z)}}{H(z)}\frac{dN_{\gamma}}{dE}(E,z) 
\times\left[\int_{0}^{r_{\rm cut}(z)} \rho_{\rm max}^{2}(z)+\int_{r_{\rm cut}(z)}^{R_{\rm UCMH}(z)}\rho^{2}(r)\right]
r^{2}dr
\label{eq:flux}
\eeqa
where $\Omega_{\rm PBH} = \rho_{\rm PBH}/\rho_{c,0}$, $H(z) = H_{0}\sqrt{\Omega_{m}(1+z)^{3}+\Omega_{\Lambda}}$. 
$\tau(E,z)$ is the optical depth and we use the results in Refs.~\cite{Bergstrom:2001jj,Primack:2000xp,Salamon:1997ac}. 
$dN_{\gamma}/dE$ is the energy spectrum 
of dark matter annihilation, which can be obtained using the public code, e.g. DarkSUSY~\footnote{https://darksusy.hepforge.org/}. 

According to the simulation, the density profile of dark matter in UCMHs 
is independent on redshift~\cite{Ricotti:2007jk,0908.0735,
Adamek:2019gns,Boucenna:2017ghj,Eroshenko:2016yve}. Therefore, 
in previous works, using the values of $\rho_{\rm max}(z=0)$ and $r_{\rm cut}(z=0)$, 
the integrated result of the second line in Eq.~(\ref{eq:flux}) is independent on redshift, and in this case 
the authors of e.g.~\cite{Boucenna:2017ghj,Adamek:2019gns} treat UCMHs to 
be analogous to the dark matter decay case. However, as mentioned in previous section, both of 
$\rho_{\rm max}$ and $r_{\rm cut}$ depend on redshift. Therefore, the integrated result 
of the second line in Eq.~(\ref{eq:flux}) also depends on redshift. In fact, with simple algebraic calculation, 
one can find that 
\beqa
\left[\int_{0}^{r_{\rm cut}(z)} \rho_{\rm max}^{2}(z)+\int_{r_{\rm cut}(z)}^{R_{\rm UCMH}(z)}\rho^{2}(r)\right]
r^{2}dr \propto M_{\rm PBH} (1+z)
\label{eq:10}
\eeqa

From Eq.~(\ref{eq:10}), it can be infered easily that for EGB the analog of UCMHs to dark matter decay 
is incomplete. Moreover, the energy spectrum $dN_{\gamma}/dE$ 
in Eq.~(\ref{eq:10}) is also different between the cases of dark matter annihilation and 
decay~\cite{Cirelli:2010xx,Cirelli:2008pk}. 
The extragalactic gamma-ray background has been observed by Fermi experiment~\cite{fermi-1,fermi-2}. 
The EGB has been used to constrain 
the properties of dark matter particle. In this work, we use the method in Ref.~\cite{fermi-3} to 
derive the conservative limits on PBHs. The basic formula is 

\beqa
\phi_{i}^{\gamma} \leq M_{i} + n \times \Sigma_{i}, 
\label{eq:fermi}
\eeqa
where $\phi_{i}^{\gamma}$ is the integrated flux in $i$th energy bin, which can be 
obtained using Eq.~(\ref{eq:flux}). 
$M_i$ and $\Sigma_i$ is the measured flux and error of $i$th energy bin, respectively. 
$n = 1.64$ corresponds to the $95\%$ confidence level. Using Eqs.~(\ref{eq:flux}) and (\ref{eq:fermi}), 
one can get the constraints on PBHs. In this work, we set the basic parameters of dark matter particle as 
$m_{\chi} =100~\rm GeV$ and $\left<\sigma v\right>=3\times 10^{-26}~\rm cm^{3}~s^{-1}$. 
We use $b\bar b$ annihilation channel for our calculations. The process of calculations can be applied easily 
to other 
parameters and channels. Using Fermi data given in Ref.~\cite{fermi-2}, we derive the constraints on the abundance 
of PBHs, $\Omega_{\rm PBH,EGB} \lesssim 2\times 10^{-8}$. From Eqs.~(\ref{eq:flux}) and (\ref{eq:fermi}), it can be 
found that the constrains on PBHs from the EGB data are independent on the masses of PBHs. In fact, as discussed 
in e.g. Refs.~\cite{Boucenna:2017ghj,Adamek:2019gns}, there should be valid mass range for these constraints and 
relevant discussions will be given in Sec. IV. 


\subsection{Constraints on PBHs from the global 21cm data}

Recently, the observational results of the EDGES experiment show an absorption feature in global 21cm signals 
at the redshift of $z\sim 17$. The amplitude of absorption trough is about twice the expected one. 
Utilizing the results of the EDGES experiment, the properties of dark matter particles can be constrained. 
The abundance of PBHs can also be constrained using 21cm data by considering Hawking radiation from PBHs, 
see e.g. Ref.~\cite{yinzhema}. 
Be different from previous works, we consider the influences of dark matter annihilation within dressed PBHs 
on global 21cm signals. 

The basic ideas of this section is to consider the interactions between the particles emitted from the dressed PBHs 
due to dark matter annihilation and 
existed in the Universe, which can affect the evolution of intergalactic medium (IGM). 
The main influences on IGM are ionization, heating and excitation
~\cite{xlc_decay,lz_decay,binyue,mnras,epjplus-2,energy_function,Slatyer:2015kla}, 
having influences on global 21cm signals. The bright temperature of global 21cm signals 
can be written as~\cite{prd-edges,binyue,Cumberbatch:2008rh,Ciardi:2003hg} 

\beqa
\delta T_{\rm 21} = ~ 26(1-x_e)\left(\frac{\Omega_{b}h}{0.02}\right)\left[\frac{1+z}{10}\frac{0.3}{\Omega_{m}}\right]^{\frac{1}{2}}\times \left(1-\frac{T_{\rm CMB}}{T_s}\right)~\rm mK.
\eeqa
where $x_e$ is the degree of the ionization of IGM. $T_s$ is the spin temperature, 
which is defined as 

\beqa
\frac{n_1}{n_0}=3\mathrm{exp}\left(-\frac{0.068{\rm K}}{T_s}\right),
\eeqa
where $n_0$ and $n_1$ are the number densities of hydrogen atoms in triplet and singlet states. 
The spin temperature $T_s$ is effected mainly by (i) the background photons; (ii) the collisions between 
the particles; (iii) the resonant scattering of $\rm Ly\alpha$ photons. 
Including these effects, $T_s$ can be written as~\cite{Kuhlen:2005cm,epjplus-2,Liszt:2001kh,binyue} 

\beqa
T_{s} = \frac{T_{\rm CMB}+(y_{\alpha}+y_{c})T_{k}}{1+y_{\alpha}+y_{c}},
\eeqa
where $y_{\alpha}$ corresponds to the resonant scattering of $\rm Ly\alpha$ photons~\cite{binyue,mnras,Kuhlen:2005cm}. 
$y_c$ corresponds to the collisions between the particles~\cite{Kuhlen:2005cm,epjplus-2,Liszt:2001kh}. $T_{\rm CMB}$ and $T_k$ are the temperature of CMB and IGM. 

As mentioned above, due to the radiation from UCMHs the evolution of $x_e$ and $T_k$ is changed. 
The equations governing the evolution of $x_e$ and $T_k$ 
can be written as~\cite{xlc_decay,lz_decay,binyue,mnras,epjplus-2,energy_function,Slatyer:2015kla}

\beqa
(1+z)\frac{dx_{e}}{dz}=\frac{1}{H(z)}\left[R_{s}(z)-I_{s}(z)-I_{\rm PBH}(z)\right],
\eeqa

\beqa
(1+z)\frac{dT_{k}}{dz}=&&\frac{8\sigma_{T}a_{R}T^{4}_{\rm CMB}}{3m_{e}cH(z)}\frac{x_{e}}{1+f_{\rm He}+x_{e}}
(T_{k}-T_{\rm CMB})\\ \nonumber
&&-\frac{2}{3k_{B}H(z)}\frac{K_{\rm PBH}}{1+f_{\rm He}+x_{e}}+T_{k}, 
\eeqa
where $R_{s}(z)$ and $I_{s}(z)$ are the recombination rate and standard ionization rate, respectively. 
$I_{\rm PBH}$ and $K_{\rm PBH}$ are the ionization and heating rate caused by 
dark matter annihilation from dressed PBHs. $I_{\rm PBH}$ and $K_{\rm PBH}$ are governed 
by the energy injection rate per unit volume $\rm dE~dV^{-1}~dt^{-1}$, and they can be written as 

\beqa
I_{\rm PBH} = f(z)\frac{1}{n_b}\frac{1}{E_0}\times
\frac{{\rm d}E}{{\rm d}V{\rm d}t}\bigg|_{\rm PBH}
\label{eq:I}
\eeqa
\beqa
K_{\rm PBH} = f(z)\frac{1}{n_b}\times \frac{{\rm d}E}{{\rm d}V{\rm d}t}\bigg|_{\rm PBH} 
\label{eq:K}
\eeqa
where $f(z)$ is the energy fraction injecting into IGM for inoization, excitation and heating. 
$f(z)$ is a function of redshift and annihilation channel. For much detailed discussions about 
$f(z)$ one can refer to e.g. Refs.~\cite{energy_function,Slatyer:2015kla,chi_1}. 
 
\begin{figure}
\epsfig{file=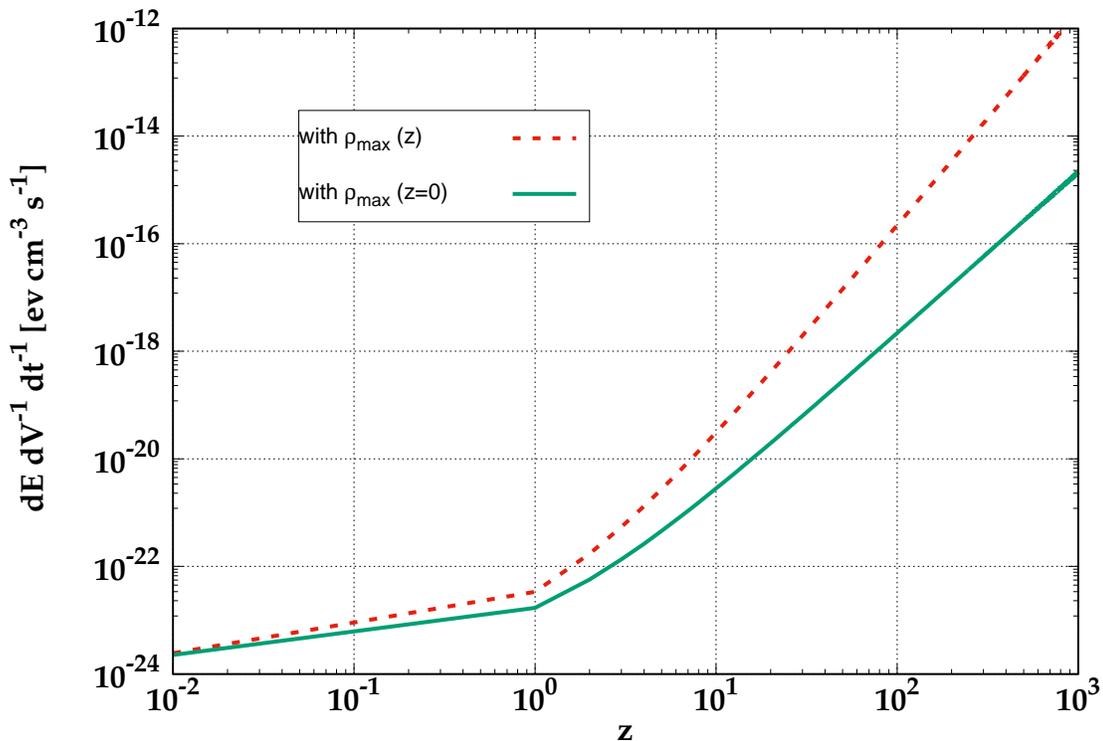,width=0.85\textwidth}
\caption{The evolution of energy injection rate per unit volume (red dashed line). 
For comparison, we also show the case of using $\rho_{\rm max}(z=0)$ for calculations (green solid line). 
In this case, the energy injection rate per unit volume is proportional to $(1+z)^{3}$. 
In this figure, we have set the parameters of dark matter as 
$m_{\chi} =100~\rm GeV$, $\left<\sigma v\right>=3\times 10^{-26}\rm cm^{3}s^{-1}$ and the abundance of PBHs as 
$\Omega_{\rm PBH} = 10^{-8}$. }
\label{fig:dedvdt}
\end{figure}

For dressed PBHs, the energy injection rate per unit volume due to dark matter annihilation can be written as

\beqa
\frac{{\rm d}E}{{\rm d}V{\rm d}t}\bigg|_{\rm PBH} = ~n_{\rm PBH}\times m_{\chi} \times 
\frac{\left<\sigma v\right>}{m_{\chi}^{2}}\int \rho^{2}(r)r^{2}dr
\propto ~ \left(\frac{\left<\sigma v\right>}{m_{\chi}}\right)^{\frac{1}{3}}\Omega_{\rm PBH} (1+z)^{4},
\label{eq:ene_inj} 
\eeqa
where $n_{\rm PBH} = \rho_{\rm PBH}/M_{\rm PBH}$, $\Omega_{\rm PBH} = \rho_{\rm PBH}/\rho_{c,0}$. 
The factor $(1+z)^{4}$ comes from the fact that the number density of PBHs is proportional to $(1+z)^3$, and 
the integrated result in Eq.~(\ref{eq:ene_inj}) is proportional to $(1+z)$ (also shown in Eq.~(\ref{eq:10})). 
The evolution of energy injection rate is shown in Fig.~\ref{fig:dedvdt}. For comparison, we have shown 
energy injection rate for $\rho_{\rm max}(z=0)$, which is proportional to $(1+z)^3$. In this work, 
we have modified the public code ExoCLASS~\cite{exoclass}, which is the branch of 
the CLASS code~\footnote{https://www.class-code.net}, 
to include the effects of dressed PBHs for getting the evolution 
of $x_e$ and $T_k$. For the parameters of dark matter particle, we have set 
$m_{\chi} =100~\rm GeV$ and $\left<\sigma v\right>=3\times 10^{-26}~\rm cm^{3}~s^{-1}$. 
We use $b\bar b$ annihilation channel for our calculations. Following previous works, by requiring 
the bight temperature of global 21cm signals $\delta T_{\rm 21} \lesssim -100~\rm mK$, 
we obtain the constraints on the abundance of PBHs, $\Omega_{\rm PBH,21cm} \lesssim 2\times 10^{-10}$. 
Be similar to the limits from the EGB data, the constraints on PBHs from 21cm data are also independent on 
the masses of PBHs.


\section{Conclusions}

We have investigated the contributions of dark matter annihilation within dressed PBHs to 
the extragalatic gamma-ray background and global 21cm signals. Many previous works have focused 
on the similar ultracompact minihalos, which would be formed directly through the collapse of 
large density perturbations existed in the early Universe. Shown by simulations, UCMHs can be formed through 
the accretion of dark matter particles onto PBHs~\cite{Adamek:2019gns,kaz}, instead of formation 
direct from the collapse of large density perturbations
~\cite{Bringmann_1,dongzhang,scott_2015,ucmh_scott,yunlongzheng,yang_2011,epjplus-2,yyp_neutrino,epl,xingyu,kaz_1,kaz_2}, 
which can be used to limit the abundance of PBHs. 

Considering dark matter annihilation, there is a maximum core density $\rho_{\rm max}$ in UCMHs, 
which has been treated independent on redshift in previous works. 
While in this work, we have made improvement and extension 
by showing $\rho_{\rm max}$ depending on redshift. Through considering 
the contributions of dark matter annihilation in dressed PBHs to the extragalactic gamma-ray background, 
we have derived the constraints on PBHs using the Fermi data, $\Omega_{\rm PBH,EGB} \lesssim 2\times 10^{-8}$. 
Dark matter annihilation in dressed PBHs can also affect the global 21cm signals. 
The recent observational results of the EDGES experiment have been used widely to investigate 
the properties of dark matter particles. In this work, by requiring the bright temperature of global 21cm signal 
$\delta T_{\rm 21} \lesssim -100~\rm mK$, we have derived the limits on the abundance of PBHs, 
$\Omega_{\rm PBH,21cm} \lesssim 2\times 10^{-10}$. 

From discussions in Sec. III, it can be seen that the constraints on PBHs are independent on their masses, 
while as shown in e.g. Ref.~\cite{Adamek:2019gns}, not valid for 
mass of $M_{\rm PBH} \lesssim 10^{-6}M_{\odot}$, where the thermal kinetic energies of 
dark matter particles can not be ignored during the process of accretion. 
In our calculations, 
we have considered the distribution of UCMHs is isotropic, ignoring UCMHs in large 
dark matter halos where the density profile of UCMHs could be effected by tidal force. Considering the 
spatial resolution of Fermi, the maximum conservative mass of UCMH here 
is about $M_{\rm PBH} \sim 10^{3} M_{\odot}$. Therefore, our limits are valid for the mass range of 
$10^{-6} M_{\odot} \lesssim M_{\rm PBH} \lesssim 10^{3} M_{\odot}$.

\section{Acknowledgements}
We thank the anonymous referees for the very useful comments and suggestions. 
This work is supported in part by the National Natural Science Foundation of China 
(under Grants No.11505005 and No.U1404114). Y. Yang is supported partly by the Youth Innovations and Talents Project of Shandong 
Provincial Colleges and Universities (Grant No. 201909118).
\

%

\end{document}